%% file: main.tex
\documentclass{cas-sc}
\usepackage[utf8]{inputenc}
\usepackage[authoryear]{natbib}
\usepackage{float}
\usepackage{wrapfig}
\usepackage{graphicx}
\usepackage{hyperref}
\usepackage{caption}
\usepackage{ulem}
\usepackage{siunitx}

\bibpunct{(}{)}{,}{n}{}{,~}
\setcitestyle{authoryear}

\newcommand{\arcsec}{$^{\prime \prime}$}

\input{Journal_Abbrevs}
\begin{document}
\let\WriteBookmarks\relax
\def\floatpagepagefraction{1}
\def\textpagefraction{.001}

\shorttitle{Dusty lunar retroreflectors}

\shortauthors{Sabhlok et al.}

\title [mode = title]{A clear case for dust obscuration of the lunar retroreflectors}

\author[1]{Sanchit Sabhlok}[orcid=0000-0002-8780-8226]
\cormark[1]
\ead{ssabhlok@ucsd.edu}

\author[3]{James B. R. Battat}[orcid=0000-0003-1236-1228]
\author[4]{Nicholas R. Colmenares}[orcid=0009-0008-6736-557X]
\author[1,5]{Daniel P. Gonzales}[orcid=0009-0008-0789-2052]
\author[2]{Thomas W. Murphy, Jr.}[orcid=0000-0003-1591-6647]

\affiliation[1]{organization={Department of Physics, University of California San Diego},
            addressline={9500 Gilman Drive},
            city={La Jolla},
            postcode={92093-0424},
            state={CA},
            country={USA}}

\affiliation[2]{organization={Department of Astronomy and Astrophysics, University of California San Diego},
            addressline={9500 Gilman Drive},
            city={La Jolla},
            postcode={92093-0424},
            state={CA},
            country={USA}}

\affiliation[3]{organization={Department of Physics and Astronomy, Wellesley College},
            addressline={106 Central Street},
            city={Wellesley},
            postcode={02481},
            state={MA},
            country={USA}}

\affiliation[4]{organization={Geodesy and Geophysics Lab, NASA Goddard Space Flight Center},
            addressline={8800 Greenbelt Rd.},
            city={Greenbelt},
            postcode={20771},
            state={MD},
            country={USA}}
\affiliation[5]{organization={Department of Physics, University of Maryland Baltimore County},
            addressline={1000 Hilltop Circle},
            city={Baltimore},
            postcode={21250},
            state={MD},
            country={USA}}

\begin{abstract}
The passive retroreflector arrays placed on the moon by Apollo 11, 14 and 15 astronauts continue to produce valuable Earth-Moon range measurements that enable high-precision tests of gravitational physics, as well as studies of geo- and selenophysics. The optical throughput of these retroreflectors has declined since their deployment, with an additional signal loss at full moon when the reflectors experience direct solar illumination. We show that the  loss in return rate can be attributed to the accumulation of a thin layer of lunar dust on the surfaces of the corner cube retroreflectors. First, a careful analysis of the optical link budget for the Apache Point Observatory Lunar Laser-ranging Operation (APOLLO) experiment reveals that the lunar return rate is 15--20 times smaller than predicted, a deficit that can be explained by a reflector dust covering fraction of ${\sim} 50$\%. Second, range measurements taken during a lunar eclipse indicate that the solar illumination of the retroreflectors degrades their throughput by an additional factor of ${\sim}15$. Finally, a numerical simulation of heat transfer in dust-coated reflectors is able to model the resulting thermal lensing effect, in which thermal gradients in the retroreflectors degrade their far-field diffraction pattern.
A comparison of this simulation to eclipse measurements finds a dust coverage fraction of ${\sim}50$\%. Taken together, the link analysis, eclipse observations and thermal modeling support the claim that the retroreflectors are obscured by lunar dust, with both link budget and simulation independently finding the dust fraction to be $\sim$50\%.
\end{abstract}



\maketitle

\section{Introduction} 
The corner cube reflectors (CCRs) placed on the moon by Apollo 11, 14 and 15 astronauts continue to produce scientific output more than 50 years after their deployment. The reflectors were designed to sit passively in the periodic temperature swings of the lunar environment in order to facilitate Lunar Laser Ranging (LLR) measurements. It would be fair to say that these reflectors have greatly outperformed expectations, continuing to operate for more than 50 years after placement. Measurements of the Earth--Moon distance provide precision tests of fundamental physics (e.g. general relativity, Lorentz Invariance, time evolution of fundamental constants), as well as geophysical information and constraints on the composition and dynamics of the lunar interior \citep{murphyLLRReview2013}.

The Apache Point Observatory Lunar Laser-ranging Operation (APOLLO) began its science campaign in 2006 \citep{murphyAPOLLO2008} with the goal of providing millimeter-accuracy range data to improve constraints on gravitational physics.

When APOLLO started ranging, it became clear that the measured return signal was lower than expected by about a factor of 10 from careful link budget calculations \citep{Murphy:2007nt}. Surprisingly, the signal fell by an additional order of magnitude when the lunar phase was within $\sim 20^\circ$ of full moon \citep{Murphy:2010CCR}.
\citet{Murphy:2010CCR}  discussed various scenarios that could lead to the degradation of performance over four decades in the lunar environment, the favored scenario being deposition of dust on the retroreflector surfaces. In a subsequent work, \citeauthor{GoodrowMurphy2012} showed that thermal gradients of a few degrees in a CCR would lead to dramatic suppression of the central intensity of the far-field diffraction pattern (FFDP) \cite{GoodrowMurphy2012}. While this does imply that the return signal from the reflectors would be lowered as a result of a thermal gradient, the paper did not perform any thermal modelling of the CCR under solar illumination. 

Lunar dust particles must rely on impacts and on electrostatic charges to levitate and transport across the lunar surface due to a lack of atmosphere \citep{Colwell2007}. Evidence for dust levitation was first seen in the western horizon pictures taken by the Surveyor 5, 6 and 7 spacecraft \citep{Criswell1973, Rennilson1974} shortly after sunset. These images show a distinct glow just above the horizon, indicating forward scattering of the incident solar flux off a levitating dust cloud $< 1$m from the ground. These images compliment the reporting of streamers originating from the lunar horizon as seen by the Apollo 17 astronauts from the spacecraft, indicating the presence of levitating dust at scale heights of 5--20 km \citep{McCoy1974, Zook1991}. Results from the Lunar Ejecta and Meteorites Experiment (LEAM) placed on the moon by the Apollo 17 astronauts indicated the presence of slow moving dust particles. These particles were not detected uniformly over the course of a lunar day, but peaked strongly around lunar sunrise, with detection rates increasing around 60 hours before sunrise and persisting 30-60 hours after sunrise \citep{Berg1976}. The origin of charge on lunar dust grains can be due to friction between meteorically-agitated dust grains, photoelectric charging from incident solar radiation or charging by the incident solar wind. The impact of solar radiation and wind on the lunar surface creates a plasma sheath over the sunlit and shadowed regions on the surface \citep{Halekas2011}. Experiments have verified that charged dust grains can levitate in such a plasma sheath, allowing for their transport across lunar surfaces \citep{Arnas2001,Sickafoose2002}. 

In this work, we present a three-pronged approach to identify the cause of reflector underperformance and lowered signal during the full moon. Firstly, we perform a detailed link budget analysis in Section~\ref{sec:link_budget}, supplemented by observations of stars of known magnitudes and by beam scans across the lunar reflectors to carefully analyze the profile of the transmitted laser beam. We then compare these calculations to observed return rates from the moon, and indeed find that the actual return rates are 15--20 times lower than expected. Secondly, in Section~\ref{sec:Observations}, we describe results of APOLLO observations during lunar eclipses---especially one in April 2014, under photometric conditions---showing that the return rate improves by more than an order of magnitude soon after a reflector enters the umbral shadow and cools. 
Section~\ref{sec:simulation} presents results from our custom thermal simulation of a CCR in which we model the expected lunar ranging return rate as a function of solar illumination during a lunar eclipse. We discuss the implications of these results in Section~\ref{sec:Discussion}, and summarize our results in Section~\ref{sec:Conclusions}.

\section{Link Budget} \label{sec:link_budget}
\input{link_budget}

\section{Eclipse Observations} \label{sec:Observations}
\input{eclipse_observations}

\section{Thermal Modelling of the Corner Cube Retroreflectors} \label{sec:simulation}
\begin{figure}
    \centering
    \includegraphics[width=\linewidth]{ 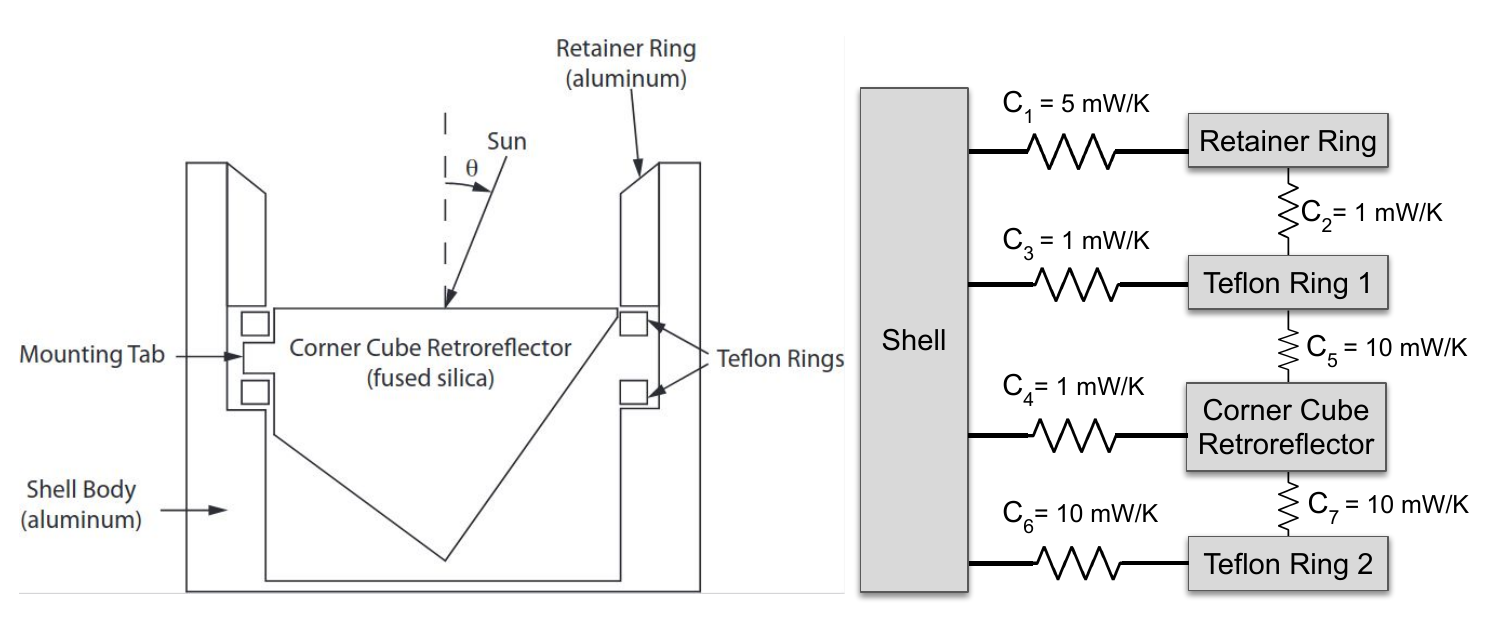}
    \caption{(Left)  Schematic diagram defining the geometry of the CCR mounted in the aluminum casing ("shell") via Teflon mounting rings and the retainer ring. (Right) Elements and conductive paths used in our thermal simulation, with specified thermal conductances. The CCR, retainer ring, and the shell radiate heat to space. Solar energy is incident on the top surfaces of the CCR, retainer ring, and shell. The corner cube looks asymmetric from this angle because the tabs are placed at 120 degree intervals around its perimeter, so diametrically opposite sides do not both have tabs.}
    \label{fig:ThermalBlockDiag}
\end{figure}
We model the optical response of a CCR under thermal load to estimate the dust coverage fraction for comparison with the degradation estimate from our link budget calculation in Section~\ref{sec:link_budget}, the eclipse observations in Section~\ref{sec:Observations}, and the known degradation of performance during full moon \citep{Murphy:2010CCR}. Our simulations were done using a custom code in C developed by our group for this purpose. 
The simulation has three stages. First, we set up the geometry of the problem, defining the different surfaces, their thermal properties and initial conditions. In the second stage, the code calculates the analytic view factors from all boundary point surface elements to all others and stores them into a view factor matrix to be referenced later for radiative heat transfer. Lastly, the code performs conductive and radiative heat transfer, based on the initial conditions for a specified amount of time. The details of the code implementation, including the geometry setup, view factor matrix construction and heat transfer, are described in Appendix \ref{app:sim_details}.

A schematic diagram and corresponding thermal model of the mounted CCR is shown in Fig.~\ref{fig:ThermalBlockDiag}. The fused silica CCR is mounted to an aluminum shell (aluminum 6061-T6), held in place by two teflon rings and an aluminum retainer ring (anodised aluminum 1100). For the CCR, retainer ring and the shell, we have to specify the net absorptivity and emissivity as a result of an impeding dust fraction sitting on the surfaces. To define the net emissivity and absorptivity, we consider a fraction $f$ of the surface to be obscured by lunar dust, with an inherent emissivity $\epsilon_{dust}$ and absorptivity $\alpha_{dust}$. For the shell and the retainer ring, the net emissivity is simply an area-weighted mean of the two emissivities. Effective emissivity and absorptivity for the top of the shell and retainer ring is given by
\begin{gather}
    \epsilon_{eff} = (1-f) \, \epsilon + f\, \epsilon_{dust} \\
    \alpha_{eff} = (1-f) \, \alpha + f\, \alpha_{dust}.
\end{gather}
The case for the top of the CCR is slightly more complicated by the fact that light can be absorbed by the dust on the way back out of the CCR as well. So the effective absorptivity comes out to be
\begin{gather}
    \alpha_{eff, CC} = [1 - (1-f)^2] \, \alpha_{dust}.
\end{gather}

Our simulation investigates both equilibrium and dynamic modes. First, a constant heat input, modelling the solar flux incident on the moon (1360 W m$^{-2}$), is used to simulate an equilibrated CCR temperature profile under full-moon illumination conditions. As our interest in this work primarily concerns observations near full moon, we do not consider the possibility of sunlight illuminating the interior of the shell behind the CCR because all incident sunlight is rejected by total internal reflection when the solar illumination is within 17 degrees of normal incidence.  The equilibrated temperature profile is then used as a starting point for an eclipse simulation, where the solar flux is varied according to an eclipse illumination profile. The simulation computes the temperature of all elements at each time step, from which the temperature profile of the CCR is extracted. 

Given the CCR temperature profile, the path length variations imposed by thermal modification of refractive index within the CCR are calculated, leading to a warped wavefront emerging from the CCR to produce a FFDP---the so-called thermal lensing effect \cite{GoodrowMurphy2012}. The effect is largely driven by the temperature gradient from the surface to the corner of the CCR. The resultant FFDP can then be compared to the case of an isothermal CCR, to quantify the loss in return rate as a percentage. We calculate this loss for the full moon illumination, which can be compared to our measured return rate during full moon observations \citep{Murphy:2010CCR}.  We also compare the modeled return rate improvement during the eclipse to that observed during the 2014 eclipse. The thermal properties for the different materials used in the simulation are outlined in Table~\ref{table:materials}.

\begin{table}[t!]
    \centering
    \caption{Properties of materials used in the thermal simulation. \label{table:materials}}
    \begin{tabular}{l l c c c c c}
    \hline
    Element & Material & Thermal Conductivity & Density & Heat Capacity & Emissivity & Absorptivity \\
    &  & $\rm{W} \, \rm{m}^{-1} \, \rm{K}^{-1}$ & $\rm{kg} \, \rm{m}^{-3}$ & $\rm{J} \, \rm{kg}^{-1} \, \rm{K}^{-1}$ & & \\
    \hline \hline
    CCR & Fused Silica & 1.38 & 2200 & 740 & 0.87 & 0.0 \\
    Rings 1 \& 2 & Teflon & 0.25 & 2200 & 1500 & 0.8 & 0.8 \\
    Shell & Aluminum 6061-T6 & 167.0 & 2700 & 896 & 0.025 & 0.06 \\
    Retainer Ring & Anodized Aluminum 1100 & 222.0 & 2710 & 904 & 0.78 & 0.32 \\
      & Lunar Dust & --- & --- & --- & 0.9 & 0.9 \\
    \hline
    \end{tabular}
\end{table}

\subsection{Steady-state Simulation Results}
To compute the steady-state thermal profile of the CCR in the presence of direct overhead solar illumination (as would be the case during a full moon), we begin the simulation with all elements set to a temperature of 300~K, and assume a solar constant value of 1360~W~m$^{-2}$. We then run the simulation until the average temperature of all four simulated elements (CCR, Teflon rings, retainer rings and shell) stabilize.

The spatial temperature profile of the CCR is extracted and used to generate a FFDP for the CCR. Different values of $f$ result in differing central maximum intensities compared to the isothermal CCR. We find that varying $f$ from 0.5 to 0.65 changes the central maximum intensities of the CCR from ${\sim}6$\% to about ${\sim}0.1$\% of the isothermal case. The peak-to-peak phase difference in the wavefront at a dust fraction of 0.5  is $\sim$ 4.9 radians due to thermal effects. This phase difference across the wavefront results in a diminished central lobe of the FFDP. The former result, at $f=0.5$, matches the full-moon deficit factor of 10--15 reported in \cite{Murphy:2010CCR}, and the results from Section~\ref{sec:Observations} and Fig.~\ref{fig:EclipseObsOverlay} of this paper.

It is worth briefly discussing the qualitative thermal behavior of the CCR when we simulate equilibrating the geometry for a constant solar flux. As the run is started, a thermal gradient is set up quite rapidly within the CCR, within the first 100 seconds of temporal evolution. The surface of the CCR absorbs heat from incident solar radiation. This creates the thermal gradient in the CCR, with the front face being hotter than the vertex. This thermal gradient is responsible for the reduced intensity in the central maximum of the FFDP. It takes far longer, a few hours, for the system to reach a steady-state temperature. The main cooling channel for the system at this point is the radiative cooling from the CCR front surface, radiating heat to space. We confirmed that the final steady-state temperatures are independent of initial conditions. Since the thermal gradients within the CCR rapidly adjust to the incident solar flux, we expect that during the eclipse simulation, the return rate should quickly improve once the reflector enters the shadow, since the gradients, rather than the absolute temperature, affect the FFDP and therefore the return rate. This is indeed the case as shown in the next subsection.

\subsection{Simulation Result Showing Eclipse Profile}\label{sec:Eclipse}
\begin{figure}[t!]
\centering
  \includegraphics[width=\textwidth]{ 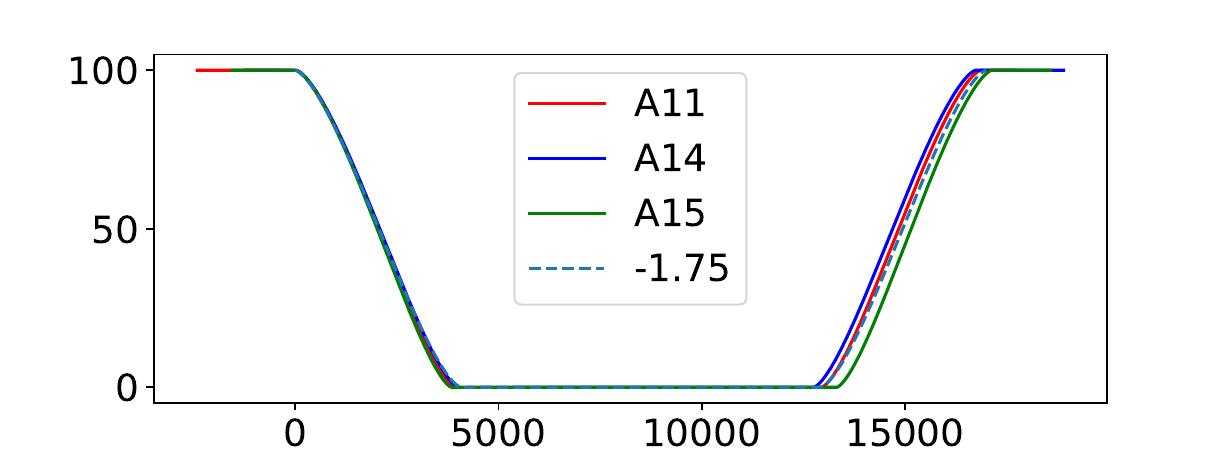}
  \caption{Percentage of maximum solar illumination as a function of time (in seconds) for the three Apollo reflectors Apollo 11 (red), 14 (blue) and 15 (green) for the eclipse on April 15th, 2014, shifted to have the same starting time. The blue dashed curve shows the eclipse profile corresponding to an impact parameter of 1.75 solar angular radii as seen from the moon, which we used in our thermal eclipse simulation since it best approximates the behavior of all three reflectors. Note that these models do not include limb darkening.} 
  \label{fig:EclipseProfile}
\end{figure}

\begin{figure}[t!]
\centering
  \includegraphics[width=\textwidth]{ 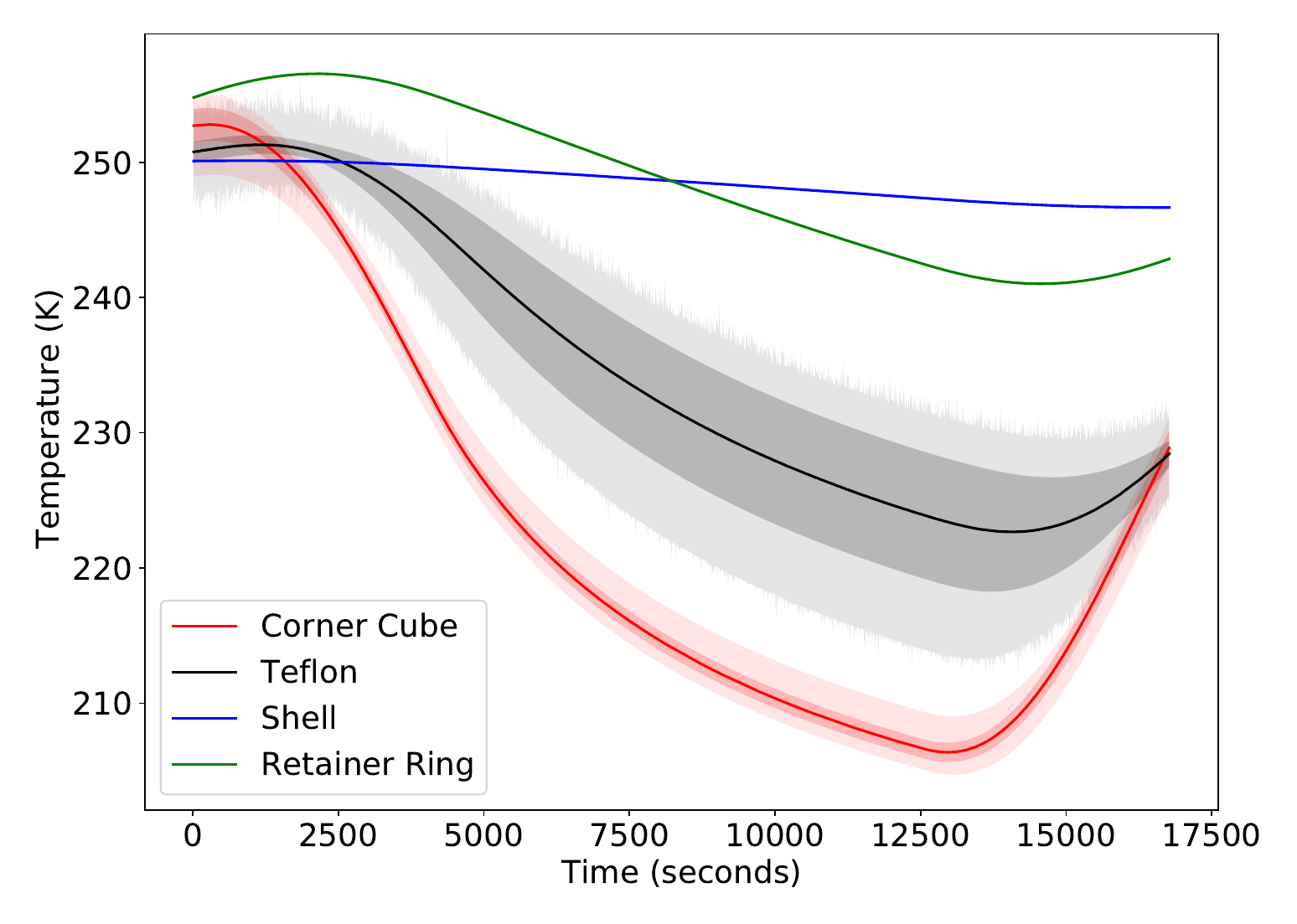}
  \caption{Average temperatures for the CCR, Teflon rings, retainer ring and  shell over the course of the simulated eclipse. Dark shading indicates $1\sigma$ deviations from the mean, whereas the lighter shading indicates the min-max. The deviations for the shell and the retainer ring are too small to be seen here due to their high thermal conductivity. The range of computed temperatures for the Teflon rings during the simulated eclipse is larger than for the other three elements. This is due to the fact that the temperatures extracted from the simulation average the two Teflon rings that are above and below the CCR. The rings also have slightly different conduction pathways to carry heat from the CCR to the shell or the retainer rings. Some of these pathways have an additional temperature dependence due to the discontinuity in thermal conductance at the Teflon-Retainer Ring and Teflon-shell interface. All of these effects combine to result in a larger range of temperatures for the Teflon rings.} 
  \label{fig:EclipseTemps}
\end{figure}

\begin{figure}[t!]
\centering
  \includegraphics[width=\textwidth]{ 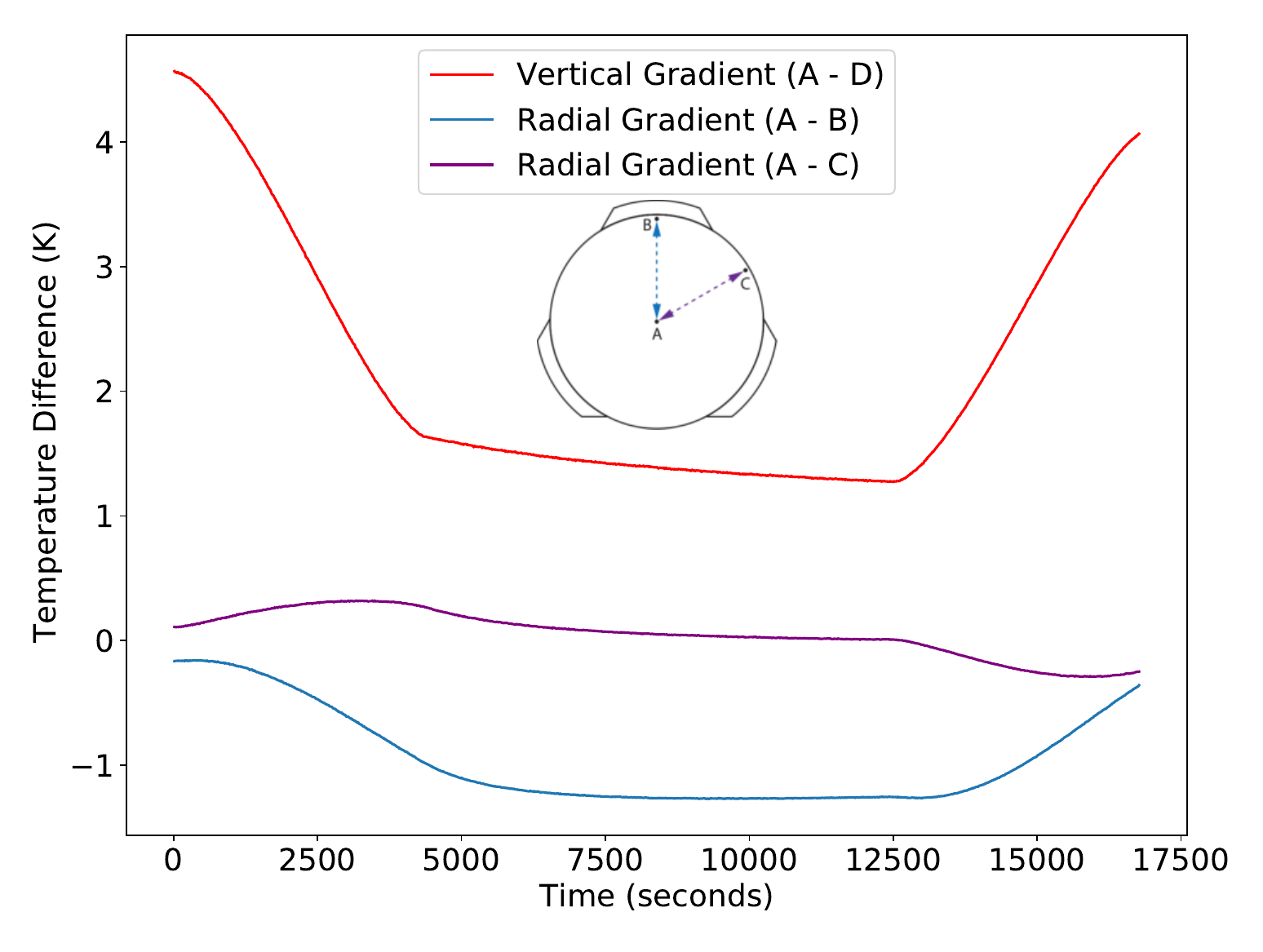}
  \caption{Temperature differences between three pairs of locations on the corner cube as a function of time during an eclipse. Corresponding solar illumination profile is provided in Fig.~\ref{fig:EclipseProfile}. As the CCR enters the umbral shadow it cools and the vertical temperature gradient drops from $\sim$5\,K to $\sim$2\,K, which reduces the thermal lensing and improves the optical throughput. The inset shows the front face of the CCR illustrating locations used for the temperature difference calculations, the center (A), tab midpoint (B) and a point between two tabs along the circumference (C). The dashed lines connecting the points correspond to the gradients shown on the plot. The vertical gradient is calculated by subtracting the temperature of the vertex (D, not shown in inset) from the center of the top surface (A - D). The radial gradient for the top surface is calculated for two points along the top surface circumference, the midpoint of the CCR tabs (A - B), and the midpoint between two tabs (A - C).  } 
  \label{fig:GradientPlot}
\end{figure}

We apply our simulation to the case of a lunar eclipse in which the solar illumination incident on the CCRs is modulated by the Earth's shadow. The relative angular sizes of the Earth and Sun as seen from the moon are used to simulate an eclipse profile for different impact parameters for the Earth's center passing in front of the Sun. This is compared to the illumination profile seen at the three different Apollo reflectors~\cite{LunarTables}. The computed illumination profile does not include limb darkening. The impact parameter that best describes the profile for all three reflectors is used as the incident solar flux profile, as shown in Fig.~\ref{fig:EclipseProfile}. While all reflectors remain under the shadow for different amounts of time, the total time difference between the shortest and longest eclipse times for different reflectors is approximately 6–7 minutes. When comparing to APOLLO observations, this is shorter than the circuit among reflectors, so the impact of different reflectors being under the shadow for different amount of times is minimized. The resultant temperatures for the CCR, teflon rings, the retainer ring and the shell are shown in Fig.~\ref{fig:EclipseTemps}. Since the FFDP evolution is largely driven by the corner cube gradients, we present the radial and vertical gradients within the CCR in Fig.~\ref{fig:GradientPlot}. For the radial gradient, due to the trifold symmetry of the corner cube, we consider a radial gradient between the center and a point on the circumference which is a midpoint of one of the ``tabs.'' We also consider a point on the circumference equidistant from the two nearest tabs for a radial gradient.

\begin{figure}[h!]
    \centering
    \includegraphics[width=\linewidth]{ 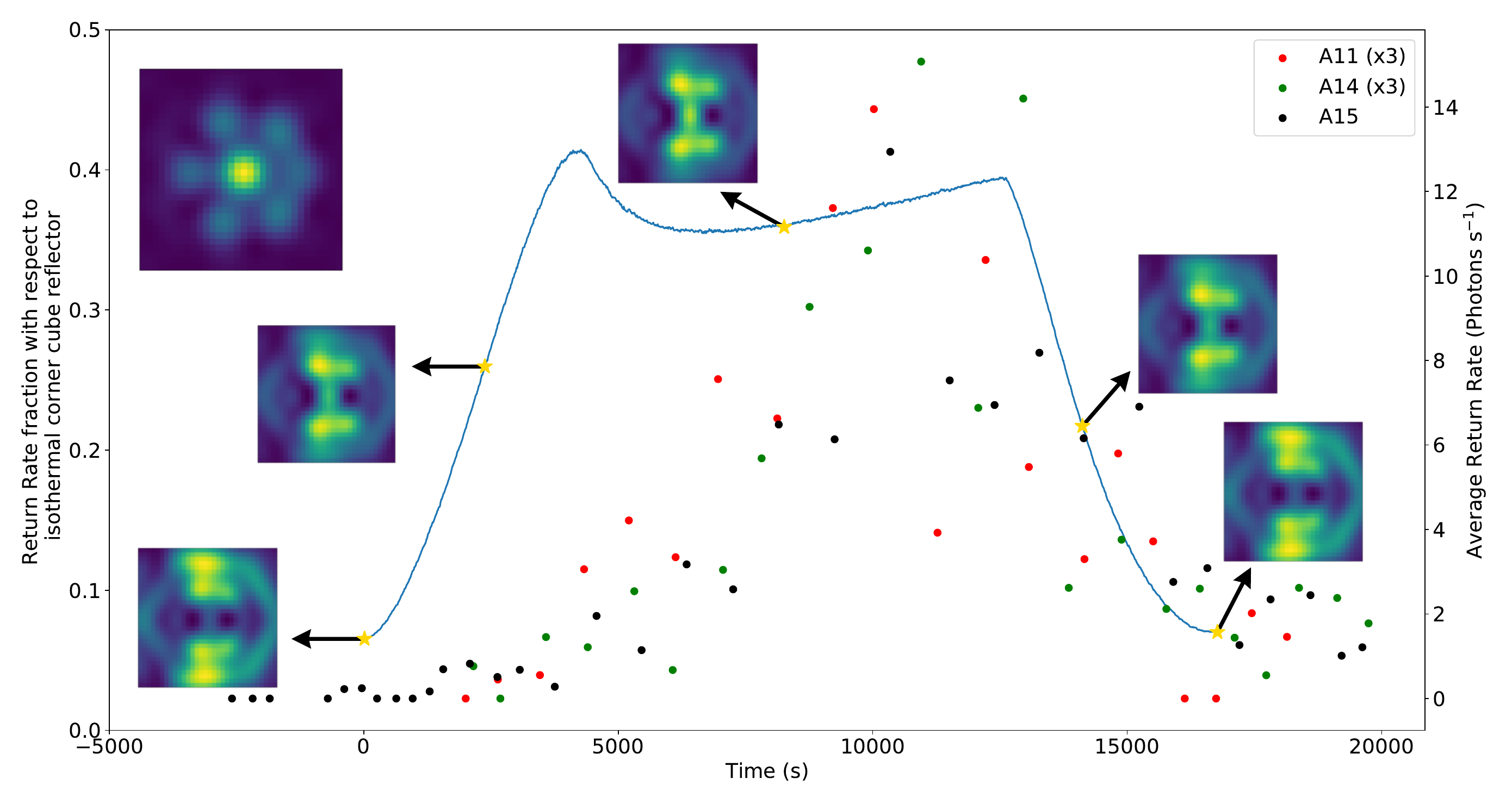}
    \caption{Fraction of light in the central diffraction maximum as compared to an isothermal CCR as a function of time during an eclipse. The measured return rates from the Apollo reflectors are also plotted (right-hand ordinate). Time is relative to when each reflector first enters the Earth shadow, which aligns all reflector time-series with the start of the eclipse simulation. Simulated CCR FFDPs are shown at five representative times during the eclipse. Inset at top left shows the FFDP for an isothermal CCR.}
    \label{fig:EclipseObsOverlay}
\end{figure}

The qualitative behavior of the system during  the eclipse can be described as follows. Starting from an equilibrated geometry, as the incident solar flux gradually decreases, the amount of heat absorbed by the CCR decreases as well. According to Figure~\ref{fig:GradientPlot}, very little radial gradient is initially present, and thus little conduction through the tabs (the chell and CCR being at a similar temperature).  Consequently, the CCR loses thermal energy primarily through radiation from the front surface. Before long, the CCR becomes cooler than the shell, so heat conducts into the CCR through the mounting tabs, setting up radial gradients while maintaining a weak vertical gradient as shown in Figures~\ref{fig:EclipseTemps} and \ref{fig:GradientPlot}. When the solar flux dwells at zero, the CCR continues to cool, but the gradient is relatively stable, thus resulting in the plateau region in the expected return rate shown in Figure~\ref{fig:EclipseObsOverlay}. When the reflectors come out of the shadow, solar flux increases so that the CCR starts to absorb more heat and the strong vertical gradient develops again, depressing the expected return rate. Throughout this process, the CCR front surface remains hotter than the vertex, i.e. the gradient maintains the same orientation.  

The time dependence of the ratio between the intensity of the central maximum in the FFDP of the CCR during an eclipse with respect to an isothermal corner cube is plotted in Fig.~\ref{fig:EclipseObsOverlay}, along with the measured return rates from individual Apollo reflectors during the 2014 eclipse. The simulated return rate increases rapidly after the CCR enters the Earth's shadow, and is correlated with a rapid reduction in the temperature gradient from the back to the front of the CCR as it cools as shown in Fig.~\ref{fig:GradientPlot}. As the CCR stays in the shadow, it retains the low vertical gradient compared to steady state illumination during full moon. When the reflectors come out of the shadow, the gradient increases to pre-eclipse levels rapidly, destroying the central maximum in the FFDP resulting in low return rates from the CCR. The simulated ratio between the maximum intensity in the FFDP during the eclipse to the start of eclipse is ${\sim}7$, as the intensity goes from ${\sim}$6\% to ${\sim}42$\%. This is comparable to the observed ratio of median return rate (0.35) during an eclipse with respect to the original rate (0.06), i.e. a factor of 6, as shown in Fig.~\ref{fig:EclipseObsOverlay} and reported in Section~\ref{sec:Eclipse}. 
Return rate observations in Fig.~\ref{fig:EclipseObsOverlay} show a low return rate prior to eclipse, followed by a rapid rise in the return rate (a factor of ${\sim}7$ in 10000 seconds) after the CCRs enter the umbral shadow. 

Our simulation includes only one CCR, and therefore will not capture all elements present in the actual reflector array on the lunar surface. However, our objective is to show that 
(1) dust reduces the intensity of the central maximum in the FFDP of the CCR and (2) as the incident solar flux on the CCR decreases during the lunar eclipse, the temperature gradient between front face and the vertex rapidly decreases, improving the return rates. Discrepancies between our simulation and the eclipse-night observations are discussed in Section~\ref{sec:Discussion}.
 

\begin{figure}[h!]
    \centering
    \includegraphics[width=\linewidth]{ 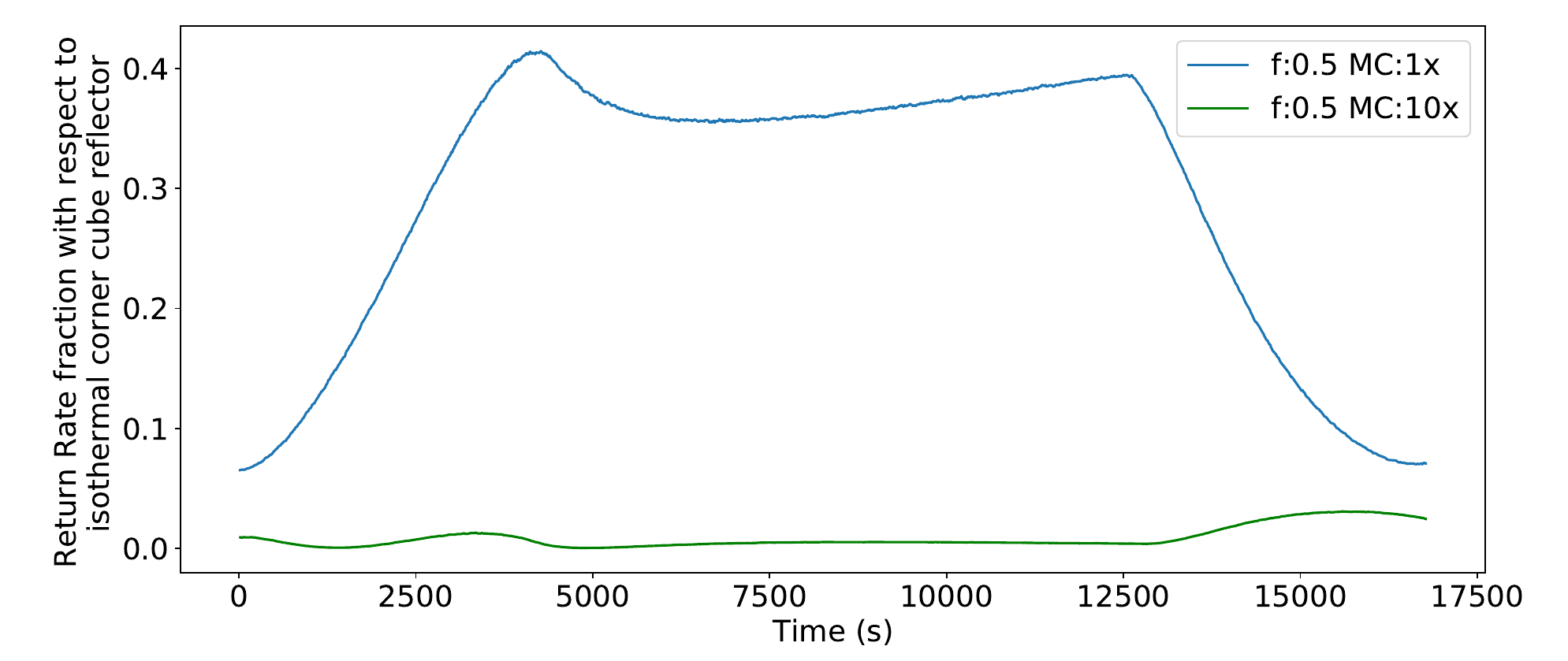}
    \caption{Simulated return rate fractions for the same dust fractions (f$=$0.5), but with the original and rescaled (x10) mount conductances (MC). The increased mount conductance almost completely depresses the intensity of the central maximum in the FFDP throughout the eclipse, and is therefore disfavored as inconsistent with observations.}
    \label{fig:MultipleEclipses}
\end{figure}

\section{Discussion}\label{sec:Discussion}
While the qualitative results from our eclipse simulation are in agreement with the measured return rate during the photometric eclipse night, there are a few details worth expanding on.

\subsection{Simulating a Single CCR}
We consider the effect of using a single CCR versus an entire array set in a pallet. We model a single dust-obscured CCR to simulate its FFDP, and compare the intensity in the central diffraction maximum to that of an isothermal CCR as an estimate of the expected relative lunar return rate. The thermal and mechanical properties implemented in this simulation have been either taken directly or derived from the description in the original Arthur D. Little Report on the CCR design and performance \cite{ADL_Report}.  However, our single cylindrical shell and CCR is not likely to track the thermal mass and environmental couplings of the real array, which could result in incorrect temporal behavior. We did not model the changing radiative temperature of the lunar surface, the geometry of coupling, or the blankets in which the arrays are wrapped. We have attempted to approximate some of these factors, but surely these are not perfect.

\subsection{The Effect of Mount Conductances}\label{sec:mountConductances}
Our goal was to model the thermal properties of the CCR using the specifications indicated in the ADL design report~\cite{ADL_Report}. A complication, however, was that the report doesn't describe the individual contact conductances but rather the total conductances through various thermal pathways. The original report uses the term "Mount Conductance" to describe the total heat conductance from outside the shell into the surface and vertex of the corner cube. To convert mount conductance given in the ADL report into contact resistances used in the simulation, we use the fact that the conductance values in ADL report are based on tests using steady state heat flux from the corner cube, through the teflon rings, through the retainer ring into the aluminum array which we have called the "shell" in our simulation. In the simulation, the conductance values for various interfaces between different materials can be tuned, however, the caveat is that the teflon-retainer ring interface is a special case. Since the retainer ring was tightened and then de-rotated by 36$^{\circ}$ at the time of assembly to prevent any thermal contact, it is likely to approximate a no-contact conductance. If this conductance value has changed since launch, it could only be due to physical contact between the two surfaces, which is likely to change conductance at this interface significantly. Thus, while our simulation can change conductance values smoothly, for the real retroreflectors, such a change will be discrete and significant. The computed conductance values for this interface were checked by calculating two extreme cases. First, we calculate the conductance for the case where there was perfect contact between teflon and the retainer ring and then we calculate the case where there is no contact between the retainer ring and the teflon ring and any thermal contact is purely radiative. Our computed conductance value lies between these two extremes, consistent with the value derived from steady state heat flux assumption based on the ADL report. We calculate that the conductance for perfect contact is approximately a factor of 10 larger than what we calculate for the APOLLO CCRs. Thus, we are satisfied that the conductance values used between the various interfaces are reasonable.

However, to study the possibility that the conductance between the teflon-retainer ring surface has increased drastically, we consider the special case of an eclipse where the mount conductance has increased by a factor of 10 as shown in Fig.~\ref{fig:MultipleEclipses}. The mount conductance values only impact our simulation if they have increased since fabrication, since a decrease would result in smaller gradients in the CCR, in agreement with our simulation and the original predictions and measurements in the ADL report. In order to run a full eclipse simulation, the modified geometry with increased mount conductances is equilibrated. The initial temperature of the shell in this case was found to be higher than the eclipse run presented in Fig.~\ref{fig:EclipseTemps}, by {$\sim$}20\,K. The expected intensity inside the central diffraction maximum is decreased approximately by a factor of 40 during the eclipse compared to the best match simulation with dust fraction of 0.5. Thus, adjusting the mount conductances completely destroys the FFDP.

Thus, given our link budget estimates and steady-state calculation of the dust coverage factor, we find that the mount conductance in our initial estimates based on the ADL report is consistent with the results from both the steady-state dust fraction calculation and the link budget analysis, and unlikely to have changed significantly from launch to present day.

\subsection{Comparisons with Infrared LLR Observations}
We can use the derived CCR temperatures for the steady state and eclipse simulations, and investigate what happens when an infrared wavelength is used ($\lambda =$ 1064 nm). For this calculation, the thermal coefficient of refractive index for fused silica is adjusted for the wavelength 1064 nm \citep{Toyoda1983}. In comparison to the steady state performance for the green laser, we find that at 1064 nm, the central intensity in the FFDP is $\sim$ 60\% compared to the case for an isothermal CCR, with a peak-to-peak phase variation of $\sim$ 2.3 radians. This marks a factor of 10 improvement in FFDP in full-moon illumination conditions for an infrared laser compared to the green laser. This agrees with improved results seen by using the infrared laser for LLR at the Grasse laser station when ranging when the lunar phase is within $\pm 18^{\circ}$ of the full moon. \citep{Grasse2017}. Note that this only includes the improved central intensity in the FFDP of the CCR, whereas observations will also depend on other factors such as detector quantum efficiency, atmospheric throughput and laser power, similar to the link budget calculation in Section \ref{sec:link_budget}.

\subsection{The Effect of Changing Dust Fraction Coverage on the CCRs}
Changing the dust fraction changes the equilibrium temperatures of the CCR, as well as the resulting gradients under solar illumination. The expected FFDP for the equilibrium temperature shows that the total intensity inside the central maximum increases as dust fraction is decreased.  A smaller dust fraction creates a smaller vertical thermal gradient in the CCR, which improves the return rate of the signal and vice versa. This return rate should become 100\% as the total dust fraction goes to zero. Given a dust fraction, the qualitative behavior of the eclipse is very similar to the one presented in Figs.~\ref{fig:EclipseTemps} and \ref{fig:MultipleEclipses}. We find that a dust fraction of 0.5 best matches the eclipse observations. 

We also note that the three reflectors Apollo 11, 14 and 15 show remarkably similar improvements during the lunar eclipse. These reflectors are placed at significant distance from each other and were placed over the course of a few years. The similarity indicates that if the dust obscuration is taken as the reason for the degrading performance, then this dust deposition must be uniformly taking place across the surface of the moon. In this case, the dust transport results from the Apollo 17 LEAM experiment \citep{Berg1976} combined with the experimental observations of dust levitation under a plasma sheath suggests a deposition mechanism \citep{Arnas2001, Sickafoose2002, Halekas2011}. The obscuration is likely caused by large particles, as indicated by the LEAM experiment, deposited due to solar wind and radiation during lunar sunrises and sunsets creating a plasma sheath, which transports these dust grains at low altitudes ($<$2\,m). Thus, our results are consistent with the lunar model of dust transport, where the transport is caused by electrostatic charges on dust particles enabling transport at low altitudes ($< 1$ m) due to the potential difference created at the time of sunrise. 

\section{Conclusions} \label{sec:Conclusions}
Measurements with the APOLLO ranging system indicate a deficit in the return rate of Apollo lunar retroreflectors by a factor of 16 to 19, with an additional order-of-magnitude loss when the reflectors are under full moon illumination. We attribute the degraded reflector performance to the accumulation of lunar dust on the reflector surface, which partially blocks lunar ranging laser light and distorts the CCR FFDP by thermal lensing due to absorbed solar radiation near full moon. This conclusion is supported by three independent studies:
\begin{enumerate}
    \item A detailed link analysis was undertaken, including a measurement of the APOLLO transmit laser beam profile, and the one-way throughput of the telescope optics and the receiver. This study found that the measured lunar return rate was only 6\% of the predicted value, corresponding to a dust fraction of ${\sim}50$\%, which in turn results in thermal consequences under solar illumination.
    \item Measurements during lunar eclipses reveal that the lunar return rate is highly dependent on solar illumination. For example, during one lunar eclipse on a photometric night, we found that the return rate improved rapidly once the reflectors entered the Earth's shadow, and  subsequently degraded when the reflectors  exit the shadow. When the reflectors are in shadow during a lunar eclipse, the median return rate is an order of magnitude higher than full-moon non-eclipse observations.
    \item We developed a heat transfer model to include conductive and radiative effects and used the simulated temperature profile of the CCR to generate the associated FFDP. We explored two scenarios:
    \begin{enumerate}
        \item Steady-state study: We ran our simulation until we achieved the equilibrium state of the CCRs  during full moon solar illumination. In the equilibrium state, the thermal gradient across the CCR degraded the intensity of the FFDP's central lobe by a factor of ${\sim}16$ compared to an isothermal CCR when a 50\% dust fraction was used, in agreement with previous assessments of the full moon deficit \cite{Murphy:2010CCR}.
        \item Time-dependent eclipse study: We then ran the equilibrated CCR setup through an eclipse simulation, roughly matching our observed eclipse conditions. Using the same dust coverage fraction of 50\%, the eclipse response improves over the full-moon return rate by a factor of 7, which compares well to the temporal profile observed during the 2014 lunar eclipse.
    \end{enumerate}
    
\end{enumerate}
Taken together, the studies above tell a consistent story:  the Apollo 11, 14 and 15 reflector surfaces are partially obscured by dust (${\sim}50$\%), which both reduces their optical throughput by the direct absorption of laser ranging photons, and through thermally induced distortions of the CCR FFDP due to the absorption of solar energy near the full moon phase.

\section*{Acknowledgements}

We thank Russet J. McMillan for her continued support as APOLLO's operator during observation, and handling of ACS functions during ranging. 
NRC’s research was supported by an appointment to the NASA Postdoctoral Program at the Goddard Space Flight Center, administered by Oak Ridge Associated Universities under contract with NASA.
This work is based on access to and observations with the Apache Point Observatory 3.5-meter telescope, which is owned and operated by the Astrophysical Research Consortium. This work was jointly funded by the National Science Foundation (PHY-0602507, PHY-1068879, PHY-1404491, PHY-1708215) and the National Aeronautics and Space Administration (NNG04GD48G, NAG81756, NNX12AE96G, NNX15AC51G, 80NSSC18K0482). JBRB acknowledges the support of the NASA Massachusetts Space Grant (NNX16AH49H).

\bibliographystyle{cas-model2-names}
\bibliography{apollo}

\appendix
\section{Thermal Simulation Details}\label{app:sim_details}
\subsection{Setting up the Geometry} \label{sub:Geometry}
Defining the geometry of the problem requires us to define a number of thermal properties for all elements of the 3 dimensional cartesian grid. The CCR are defined in detail in ADL Inc. report \citep{ADL_Report} and for the purposes of the simulation, we consider the properties defined therein to be a starting point for our simulations. We have a grid of size $51\times 51\times 51$ (102 mm $\times$ 102 mm) over which these properties are defined.

While the code as written can be generalized to an arbitrarily shaped cuboid, we work here with cubes for simplicity in view factor calculations later. While we have run the code at different spatial resolutions for testing purposes, all runs for science purposes have been run at a resolution of 2\,mm, which makes our simulated box size $102\times102\times102$\,mm. Since the heat transfer is an explicit finite difference scheme, given the material properties used and for ease of comparison for different runs, we have fixed the temporal resolution at 0.005 seconds for all runs. Since the simulation uses cubic voxels, we have oriented the CCR along the (1,1,1) direction, in order to create smooth planes for the sides of the CCRs. While this results in voxelization of the planar surfaces, the effects on heat transfer are minimal. There are two kinds of numerical effects that might occur as a result of voxelization of a planar surface of the CCR. Firstly, the amount of radiative heat transferred between two elements may be slightly inflated due to the increase in surface area as a result of the voxelization. This effect decreases with a decrease in element size, theoretically going to zero as the size of the voxel becomes infinitesimally small. We tested our setup by creating two parallel slabs oriented along the (1,1,1) direction and prescribed a heat flux on one side of one of the slabs, which then radiatively transferred heat to the other slab. Since this is a problem with an analytic solution, we were able to compare our numerical results with the analytic solution and found them to show reasonable agreement with each other. The second effect we need to be careful about is the amount of incident solar flux illuminating each voxel. We tackled this problem by analytically prescribing the heat flux for the entire voxel by multiplying the incident solar flux with the projected area along the (1,1,1) direction.

We first address the assumptions required to simulate a single CCR instead of an entire retroreflector array. The retroreflector array housing the CCRs on the moon is machined out of Aluminum 1100. One of the primary goals of the design of the reflector array was to provide passive thermal control, since it was understood any thermal gradients would have a significant impact on reflector performance. Thus, the array as a whole was thermally isolated from the lunar environment with multiple blankets, reducing external emissivity to 0.01. Additionally, the CCRs are mounted on Teflon rings and the cavities are recessed into the array to prevent illumination from sunlight unless directly overhead. These designs work well enough that we can treat an individual CCR as a standalone retroreflector for thermal purposes, as we will demonstrate with our simulation. As a result, we have simply replaced the array with an outer layer of aluminum 1100 which we refer to us as the "Shell" in this paper. As we will show, the absence of the remainder of the array has no impact on the thermal performance of the CCR. Thus, our geometry consists of a fused silica CCR, two Teflon rings, the upper aluminum retainer ring and the outer shell. The material properties for these are described in Table~\ref{table:materials}. 

The heat transfer is an explicit finite difference scheme. We first define the array $K_{i,j,k}$, which is total conductance between adjacent cells. So for a cell with indices (i, j, k) and a side length $\rm{\Delta}x$, with thermal conductivity denoted by $\rm{\lambda}_{i,j,k}$, the conductance with respect to its adjacent cell (i, j, k+1) can be calculated as -
\begin{equation}
    K_{i,j,k+1/2} = \frac{{\Delta x}^2}{\frac{\Delta x}{2 k_{i,j,k}} + \frac{\Delta x}{2 k_{i,j,k+1}} + R_{i,j,k+1/2}}
\end{equation}
Where $R_{i,j,k+1/2}$ is the contact resistance between adjacent cells. This is zero for adjacent cells with identical thermal conductivity, i.e. in the bulk of the material. For elements with distinct values of thermal conductivity, we prescribe the value of contact resistance based on the total conductance allowed between the respective surfaces. Based on the results of ADL report, the CCRs were designed to have a prescribed amount of total conductance between materials and the testing of the CCRs tried to constrain the total conductance value. For our purposes, we took the values of total conductance given in the ADL report as starting values. For pairs of material where no value was given, we prescribed the total conductance as follows. We calculated the total conductance in two limiting cases. Firstly, when the two adjacent cells have perfect contact and the total conductance comes purely from the difference in thermal conductivity of the two cells, we can set the value of contact resistance to zero. This gives a minimum conductance value. In the other limit, we assume the two contact surfaces are separated by an infinitesimal distance and the "conductance" is a result of only radiative transfer between the two surfaces. This gives us a temperature dependent total conductance value, assuming the temperature difference between the two surfaces is small compared to the absolute temperature. We then calculate this conductance for the expected temperatures on the lunar surface and get an upper limit for the total conductance. For the actual simulation runs, we end up taking the average of these two conductance values as a starting point. Once we have prescribed the total conductance for a pair of materials, we can then calculate the conductance per element. Knowing the thermal conductivity values, we can then prescribe a value of contact resistance for each element. Thus, prescribing the total conductance between any pair of materials, as designed and tested in the ADL report, gives us a value for the contact resistance between any two elements.

\subsection{Constructing the View Factor matrix}
Once we have defined our geometry, the code then constructs a view factor matrix. We create a matrix of all boundary point elements and compute the view factors for each pair of surfaces using the analytic expressions from \citeauthor{ViewFactorEhlert1993}. Since our geometry does not change as a function of time, this is a one time calculation performed at the beginning of a run. It can then be saved and retrieved for a subsequent run. This step of the code takes about 20 minutes to run for the full CCR simulation. We checked our view factor calculations for parallel and plane geometry by constructing two slabs in a parallel and perpendicular configuration. We calculate the view factor matrix, and using the view factor matrix, sum the individual view factors to obtain the total view factor for the slabs, which can then be computed using the same expression for a larger slab. We can also check this against simpler expressions for the special case where the slabs are exactly opposite each other with no relative lateral displacement. We find that the view factors computed are consistent when summed with what we would expect for the pair of larger slabs. We also find the consistency does not depend on the spatial resolution of the geometry, as would be expected for an analytic expression for the view factors. 

\subsection{Heat Transfer}
Given the conductance between adjacent cells, we can calculate the total heat influx into a cell $H_{i,j,k}$ from
\begin{align}
\begin{split}
    H_{i,j,k} = &K_{i-1/2,j,k} \cdot (T_{i-1,j,k} - T_{i,j,k}) + K_{i+1/2,j,k} \cdot (T_{i+1,j,k} - T_{i,j,k})\\
    + &K_{i,j-1/2,k} \cdot (T_{i,j-1,k} - T_{i,j,k}) + K_{i,j+1/2,k} \cdot (T_{i,j+1,k} - T_{i,j,k})\\
    + &K_{i,j,k-1/2} \cdot (T_{i,j,k-1} - T_{i,j,k}) + K_{i,j,k+1/2} \cdot (T_{i,j,k+1} - T_{i,j,k}).
\end{split}
\end{align}
The change in temperature $\rm{T}$ in time $\rm{\Delta}t$ is given by
\begin{equation}
    T^{t+1}_{ijk} = T^{t}_{ijk} + H_{ijk} \frac{\Delta t}{{\Delta x}^3 \rho_{i,j,k} C_{i,j,k}},
\end{equation}
where $\rho_{i,j,k}$ is the density and $C_{i,j,k}$ is the Specific Heat Capacity for the element $(i,j,k)$. For every boundary point, we also track the temperature at the center of the external surface. 
In order to check our heat transfer code, we performed a number of tests. Our general goal was to run simulations with analytic solutions in order to have something concrete to compare to. To test the conduction code, we use a single slab and subject it to an incident flux on one side. We then switch off any radiative effects on the boundaries, thus reducing this to a one-dimensional heat transfer problem with a known solution. We can start the simulation run from an arbitrary temperature and let it run until we have equilibrium. In practice, we started approximately $50^{\circ}$ away from the steady state temperature value. We compared the temperatures on the front and back surfaces with the analytically calculated expected temperatures. We found our simulation agreed with the analytic solution to well within $0.5 \%$ for the coarsest resolution of 1 cm and to well within $0.1 \%$ using the finest resolution of 1 mm. We then tested our radiative heat transfer code by adding a second slab in parallel behind the back face of the first slab, with boundary radiative effects switched off. We had an incident flux on one side of the first slab. Now we can analytically compute 4 temperatures on the front and back surfaces of the two slabs as follows. Assuming the temperatures on the front and back of the first slab are $T_1$ and $T_2$, and that on the front and back of the second slab are $T_3$ and $T_4$, the relevant equations in steady state are
\begin{gather}
Q - \sigma T_1^4 = - k (T_1 - T_2) \\
- k (T_1 - T_2) = \sigma T_2^4 - F_{23} \sigma T_3^4 \\
F_{23} \sigma T_2^4 - \sigma T_3^4 = - k (T_3 - T_4) \\
- k (T_3 - T_4) = \sigma T_4^4,
\end{gather}
where $k$ is the thermal conductivity of the slab, $\sigma$ is the Stefan-Boltzmann constant, $Q$ is the incident heat flux per incident area on the front surface and $F_{23}$ is the view factor between the back and front surfaces of slabs 1 and 2 respectively. Solving these sets of equations numerically gives only two real sets of values, only one of which is physical (positive temperatures).
We calculated these values and found good agreement between the numerically obtained and the solutions to the analytic equations, with the agreement improving with higher numerical resolution. 

\end{document}

%% file: link_budget.tex
The laser used for APOLLO generates upwards of $10^{17}$ photons per pulse.  Under good conditions, only about one round-trip photon is detected per pulse.  A multitude of factors---dominated by divergence---contribute to such a high attrition rate.  This section evaluates the link budget, aiming to account for all signal losses along the beam's round-trip journey from the laser to the detector by way of the moon.  We will compare the results of the link equation, Eq.~\ref{eq:link}, to observations on two high-performance epochs, adjusting some of the parameters to suit changing conditions. The nominal link budget is:

\begin{equation}\label{eq:link}    N_\mathrm{detect}=N_\mathrm{launch}\eta_\mathrm{launch}\eta_\mathrm{c}^2\eta_\mathrm{r}\eta_\mathrm{NB}Q\eta_\mathrm{FOV}\eta_\mathrm{refl}N_\mathrm{refl}p_\mathrm{gauss}\bigg(\frac{d}{r\phi}\bigg)^2p_\mathrm{refl}\bigg(\frac{D_\mathrm{eff}}{r\Phi}\bigg)^2 \prod_{i}D_i.
\end{equation}
Here, the number of photons we expect to detect is denoted as $N_\mathrm{detect}$.  Table~\ref{tab:link-params} elucidates the meaning of the terms on the right, elaborated in the text. 

\begin{table}[tbh]
    \begin{center}
    \caption{Parameters used in link equation (Eqn.~\ref{eq:link}).  Values for $\eta_\mathrm{r}$ and $Q$ are not determined separately, using the measured amalgam $\eta_\mathrm{c}\eta_\mathrm{r}\eta_\mathrm{NB}Q$ instead.  Blank uncertainties are either meaningless or too small to matter.  Some parameters are adjusted for specific observations.}\label{tab:link-params}
    \begin{tabular}{llcc}
    \hline
    Parameter             & Description                      & Value                & Uncertainty \\
    \hline
    $N_{\mathrm{launch}}$ & Laser photons per pulse          & $2.7\times 10^{17}$  & $0.14\times 10^{17}$ \\
    $\eta_\mathrm{launch}$& Central obscuration in uplink    & 0.60                 & 0.03                 \\
    $\eta_\mathrm{c}$     & Common-path optical throughput   & 0.53                 & 0.09                 \\
    $\eta_\mathrm{r}$     & Receiver efficiency              & ---                  & ---                  \\
    $\eta_{\mathrm{NB}}$  & Narrow-band filter transmission  & 0.99                 & 0.003                \\
    $Q$                   & Photon detection efficiency      & ---                  & ---                  \\
    $\eta_\mathrm{c}\eta_\mathrm{r}\eta_\mathrm{NB}Q$ & One-way throughput amalgam & 0.053     & 0.007                \\
    $\eta_\mathrm{FOV}$   & Limited detector field of view   & 0.787                & adjusted             \\
    $\eta_{\mathrm{refl}}$& Corner cube optical transmission & 0.93                 & 0.005                \\
    $N_{\mathrm{refl}}$   & Number of reflectors in array    & 100 or 300           & ---                  \\
    $p_{\mathrm{gauss}}$  & Gaussian profile vs. top-hat     & 0.693                & ---                  \\
    $d$                   & Diameter of single corner cube   & 0.0381 m             & ---                  \\
    $r$                   & Earth-Moon distance              & $3.85\times 10^8$ m  & adjusted             \\
    $\phi$                & Uplink beam divergence           & 1.0 \arcsec           & adjusted             \\
    $p_{\mathrm{refl}}$   & TIR CCR diffraction vs. top-hat  & 0.182                & ---                  \\
    $D_{\mathrm{eff}}$    & Effective telescope aperture     & 3.26 m               & ---                  \\
    $\Phi$                & Downlink beam divergence         & 2.89 \arcsec          & ---                  \\
    $D_{i}$                & Adjustments for epoch particulars & Table \ref{tab:derating} & ---          \\
    \hline
    \end{tabular}
    \end{center}
\end{table}

APOLLO's laser operates at a wavelength $\lambda=532$~nm, and the energy per pulse of the laser is $\sim 0.1\mathrm{\,J}$, resulting in $N_\mathrm{launch}\approx 2.7\times10^{17}$ generated photons per pulse.  Transmission losses are accounted later, but we separately address the geometrical loss of the outgoing Gaussian beam due to the secondary mirror obstruction via $\eta_\mathrm{launch}=0.60\pm 0.03$.

Various transmission and reflective efficiencies associated with specific segments of the photons' overall path are captured by $\eta$ factors in Eq.~\ref{eq:link}.  The first efficiency, $\eta_\mathrm{c}$, is the one-way optical efficiency that is common to both the transmit and receive paths of the APOLLO system---including atmospheric losses.  For Apache Point, $\eta_c$ evaluates to $0.53\pm0.09$ at 532~nm under clear skies and pointing at zenith (three reflections at $0.85\pm 0.05$ and 0.87 atmospheric transmission). Because this path is traversed twice---once for the transmitted beam and once for the received signal---$\eta_\mathrm{c}$ appears squared in Eq.~\ref{eq:link}.  The combined term $\eta_\mathrm{c}\eta_\mathrm{r}\eta_\mathrm{NB}Q\approx 0.053$ relates to the total one-way throughput of the receiver/detector, which is experimentally determined as a group in Section~\ref{sec:stare}.  The $\eta_\mathrm{FOV}$ term, nominally computed for a 1.0~arcsec seeing disk, represents the fraction of the point spread function (PSF) captured by the 1.4~arcsec square field of view of the APOLLO receiver, with one corner pixel in the $4\times 4$ array detector unused.

The corner cube reflectors (CCRs) number $N_\mathrm{refl}$ in the reflector assembly (100 for Apollo~11 and 14; 300 for Apollo~15), have a front-surface reflection loss represented by $\eta_\mathrm{refl}$, and diameter, $d$.  

The first squared term in parentheses represents the fraction of the uplink beam reaching a single CCR, according to a top-hat profile having a nominal diameter of $\phi=1.0$~arcesc (see Section~\ref{sec:beam}).  For the divergence geometry, the one-way telescope-reflector distance is set to the average Earth--Moon distance of $r=3.85\times 10^8$~m.  We later correct for the geometry at the epoch of observation via derating terms, $D_{i}$.  The erroneous top-hat model is then corrected to describe a Gaussian profile more characteristic of atmospheric seeing, whose full-width at half-maximum (FWHM) matches the diameter of the top-hat profile.  The correction factor, $p_\mathrm{gauss}=\ln 2$, accounts for the reduced central intensity of a Gaussian profile having the same total flux and FWHM as the top-hat (FWHM is $\sqrt{\ln 256}$ times the Gaussian sigma).  

The second squared term in parentheses captures geometric loss in the returning beam, whose divergence is also initially treated as a top-hat illumination profile of angular diameter $\Phi=\lambda/d$. This is intercepted by a telescope whose effective diameter is such that the collecting area is $\pi D^2_{\mathrm{eff}}/4$.  The top-hat model is then corrected to represent the central intensity of a far-field diffraction pattern arising from total internal reflection (TIR) by the parameter $p_\mathrm{refl}$ \cite{MurphyGoodrow2013}.

Note that while not appearing explicitly in Table~\ref{tab:link-params}, a characteristic uncertainty of 0.05~arcsec is applied for $\phi$, which propagates into an uncertainty value for $\eta_\mathrm{FOV}$ of 0.03.  These uncertainties are applied with the rest in Section~\ref{sec:link-assessment}.

\subsection{Beam Scan Test: Lunar Beam Profile}\label{sec:beam}

In order to assess the uplink divergence, $\phi$, we contrive to scan the telescope-fixed laser across the reflector while keeping the detector centered on the reflector.  This requires a coordinated move of the receiver's adjustable offset pointing angle to counter the motion of the telescope (and thus outgoing beam).  We perform a scan in two dimensions, effectively in a plus-shaped pattern centered on the nominal pointing.  As each pixel in the $4\times 4$ detector only spans 0.35~arcsec, very small motions are noticeable.

As such, a scan of this sort is easily thwarted by pointing drift.  Moreover, even when maintaining the pointing to the best of our ability, factor-of-two variations in signal on the timescale of tens of seconds are routine, largely due to seeing variations (APOLLO is highly sensitive to seeing, both due to target illumination and overfilling of the small detector).  This makes interpreting a beam scan tricky: how much of the variation is due to the pointing offset, how much is variable seeing, and how much is due to variation in other terms in the link budget?

To mitigate this, we execute a pattern of moves that returns to the central position after every offset, thus interleaving central positions between each excursion.  During the times pointing at the nominal (central) position, we sometimes adjust the pointing to re-center and maximize signal.  Furthermore, we only consider the signal change in the several seconds surrounding the transition from one position to the next, rather than averaging the whole offset result.  Each pointing therefore gives two signal ratios: at the beginning of the move, and again at the end.

\begin{figure}
\centering
\includegraphics[width=10cm]{ 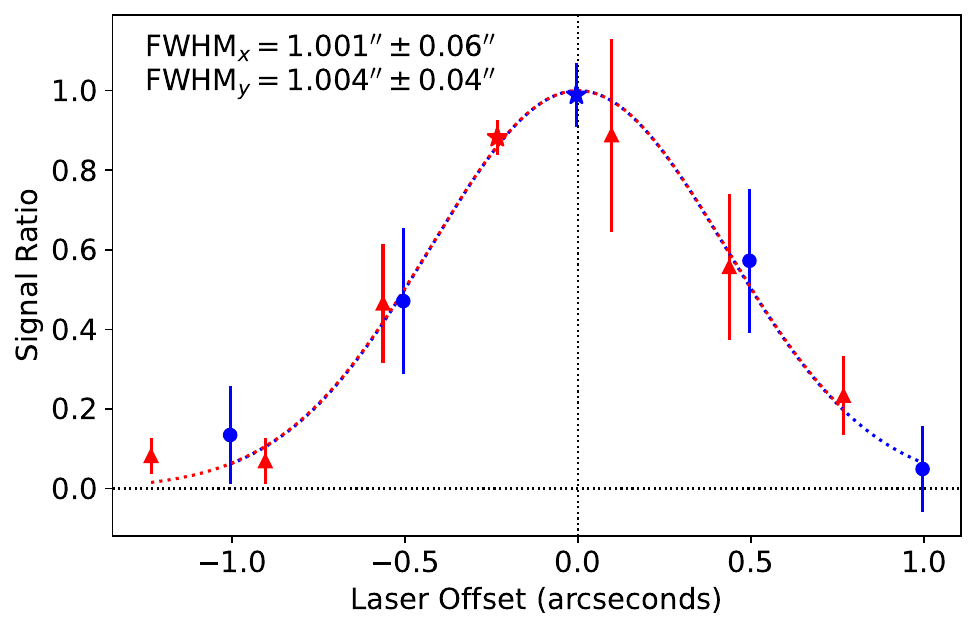}
\caption{Beam scan results on 2018 February 25 for $x$ (blue circles) and $y$ (red triangles) directions, normalized to the peak.  The nominal center positions are denoted by stars.  In this case, the FWHM of the Gaussian fits are $1.001\pm 0.06$ and $1.004\pm 0.04$ arcseconds.\label{fig:beam-scan}}
\end{figure}

Fig.~\ref{fig:beam-scan} shows an example from one epoch, in orthogonal dimensions.  Each data point averages the beginning and end ratios for the excursion.  Since all measurements are in relation to the return rate at the central position,  the result is normalized to the (changing) central value for each paired measurement.  

We report here results from beam scans on two epochs.  For 2018 February 25, we measure an uplink beam FWHM of $1.00\pm 0.05$~arcsec, while on 2018 September 28 we get $1.07\pm 0.19$~arcsec.  Because the $\phi$ parameter in Eq.~\ref{eq:link} is set to 1.0~arcsec, and squared in the link equation, we apply a derating factor of 0.88 to the latter case (the former needing no correction), as will be summarized in Section~\ref{sec:derating}.  The distribution of lunar return photons on the receiver array further corroborates---and closely matches---the beam scan results by providing an in-situ characterization of the seeing at the time of observation.

\subsection{One-Way Throughput}\label{sec:stare}

A number of the terms in Eq.~\ref{eq:link} can be simultaneously determined by pointing to a non-variable bright star of known magnitude and spectral type, measuring the photon counts, then comparing to theoretical expectations.  We first assess the top-of-atmosphere flux density at 532~nm (in W~m$^{-2}$~nm$^{-1}$, e.g.) for the spectral type and magnitude, correct for the zenith angle, sum over the effective telescope aperture, multiply by the 2.1~nm bandwidth of the narrow-band filter, and by the effective integration time (typically 500 exposures at 100~ns each), and divide by photon energy to arrive at the number of photons we would detect in the case of no losses.  By comparing actual counts, we effectively measure the product $\eta_\mathrm{c}\eta_\mathrm{r}\eta_{\mathrm{NB}}Q\eta_\mathrm{FOV}$.

In practice, we fit a two-dimensional symmetric Gaussian profile to each 50~$\mu$s exposure (typically delivering about 1,000 counts across 15 active elements) to determine the theoretical total flux: $2\pi A\sigma^2$, where $A$ is Gaussian amplitude and $\sigma$ is the standard deviation width parameter---if unconstrained by $\eta_\mathrm{FOV}$.  This produces a more stable flux estimate than the variable, jumpy seeing produces in raw terms---since the fraction of light captured in the detector is sensitive to offset and width of the PSF in a way that the total flux is not.

Several such tests on 2019 July 20 under the same detector configuration as for the beam scan epochs resulted in a determination for $\eta_\mathrm{c}\eta_\mathrm{r}\eta_\mathrm{NB}Q$ of $0.053\pm 0.007$.  We need only multiply by another factor of $\eta_\mathrm{c}$ to represent a cluster of terms in Eq.~\ref{eq:link}.  The term $\eta_\mathrm{FOV}$ can be computed exactly for a centered Gaussian in a $4\times 4$ grid whose pixels are 0.35~arcsec and missing one corner.  For example, a Gaussian whose FWHM is 1.0~arcsec delivers 78.7\% of its flux to such an array, while 1.5~arcsec seeing would result in 50.6\%.  We use the 1.0~arcsec value in Eq.~\ref{eq:link}, then correct for conditions in Section~\ref{sec:derating}.

\subsection{Derating Factors}\label{sec:derating}

Some of the parameters in Eq.~\ref{eq:link} are not constant, and require adjustment for the time of observation.  Moreover, a few additional factors impact the link and need to be considered.  These terms appear at the end of Eq.~\ref{eq:link} as a product of derating factors.  The various derating factors applied in this analysis appear in Table~\ref{tab:derating}.

\begin{table}
  \caption{Link budget derating factors for the two beam profile experiments.
  }\label{tab:derating}
  \begin{center}
  \begin{tabular}{ l|cc|cc } 
\hline
 Cause & Epoch A Value & Derating & Epoch B Value & Derating \\
    & 2018 Feb 25 & & 2018 Sep 28 & \\
\hline
  Range ($r$)         & $3.6008 \times 10^{8}$ m & 1.31  & $3.7724\times 10^{8}$ m & 1.08  \\
  Zenith Angle        & $19.9^{\circ}$           & 0.98  & $47.8^{\circ}$          & 0.87  \\
  Angular Offset (libration) & $5.7^{\circ}$            & 0.75  & $8.13^{\circ}$          & 0.66  \\
  Sun Angle ($\theta$)& $-67.3^{\circ}$          & 0.84  & $40.8^{\circ}$          & 0.74  \\
  Velocity Aberration & 0.946 \arcsec             & 0.762 & 1.003 \arcsec            & 0.736 \\
  Uplink Beam ($\phi$)& 1.00 \arcsec              & 1.0   & 1.066 \arcsec            & 0.88  \\
  Detector Capture ($\eta_\mathrm{FOV}$) & 0.90 \arcsec & 1.05 & 0.91 \arcsec        & 0.950 \\
\hline
  Total Derating & & 0.648 & & 0.282 \\
\hline
\end{tabular}
\end{center}
\end{table}

First, the Earth--Moon distance varies by as much as 13\% (full range), and appears to the fourth power in Eq.~\ref{eq:link}.  Based on the distance, $r_\mathrm{obs}$, at the time of observation, we apply a ``derating" factor (that indeed enhances the link result as often as not) of $(r/r_\mathrm{obs})^4$.

The second factor accounts for the fact that we do not operate at zenith pointing, and is provided by the expression $0.87^{2(\sec z -1)}$. Here, $z$ is the zenith angle of the Moon.  The base of 0.87 comes from the atmospheric loss by passing through one atmosphere at 532~nm.  The factor of two in the exponent is to account for the beam passing through the atmosphere twice.  Finally, we subtract one in the exponent because we already account for one atmosphere of thickness in $\eta_\mathrm{c}$.

Three other derating factors come from the geometry of the Earth--Moon and Moon--Sun relationships.  First, libration of the moon results in an angular offset of the CCR array relative to the Earth--Moon line, producing a reduction of the central irradiance of the far-field diffraction pattern via reduced area and increased diffractive spread.  Next, thermal computations by the reflector manufacturer \cite{A15Report1972} anticipated modest dependence of reflector behavior on sun angle $(\theta) $.  Finally, velocity aberration shifts the return far-field diffraction pattern on the surface of the Earth by approximately one arcsecond, so that the telescope sits not at the peak of the pattern, but on the dimmer shoulder.  The central lobe of the complex TIR diffraction pattern closely follows that of an Airy pattern from a circular aperture of the CCR's diameter \cite{MurphyGoodrow2013}.  The derating amount is thereby computed using this profile based on the relative tangential velocity at the time of observation---dominated by Earth rotation.  

Lastly, atmospheric seeing impacts both the outgoing beam divergence ($\phi$), and the fraction of light captured by the detector, $\eta_\mathrm{FOV}$.    The latter is calculated according to a Gaussian fit to the lunar return distribution on the $4\times 4$ detector, whose FWHM is given in Table~\ref{tab:derating}, along with the corresponding adjustment relative to the nominal case presented in Table~\ref{tab:link-params}.  It is reassuring that the FWHM values for the beam scans are not much larger than the FWHM values of the lunar signal on the detector---indicating good beam collimation so that the outgoing beam is essentially seeing-limited.  Note that on 2018 September 28, an additional detector element near the center was inoperative, which is why the derating factors are different, despite similar FWHM measures for the two cases.

\subsection{Link Assessment and Implied Dust Fraction}\label{sec:link-assessment}

The previous sections laid the groundwork for comparing observed lunar return rates to theoretical expectations.  Application of Eq.~\ref{eq:link} using values in Table~\ref{tab:link-params} and Section~\ref{sec:stare} produces for Apollo~15 a nominal expectation of $19.1\pm 4.8$ photons per shot, before derating factors are applied.  For the two epochs considered, the derating factors yield approximately 12.4 and 5.4 photons per shot.

On these two nights---the ones for which beam scans were performed, thus obtaining confident optimization of the signal---we assess the photon return rate at the many visits to the central beam position.  Because the LLR signal rate is extremely sensitive to telescope pointing and atmospheric seeing (via $\phi$ and $\eta_\mathrm{FOV}$ parameters), we see much variation in central-position rates throughout the roughly ten-minute scan procedure.  We therefore take special care in assessing a ``reasonable" return rate, looking to identify a consistent ``best" central rate that recurs multiple times throughout the scan and that is therefore compatible with the beam divergence assessed over the entire scan period.  We believe these ``best" rates are more representative of the theoretical link performance of APOLLO, for more meaningful comparison to the link calculation.

The estimated return rates for the two epochs are assessed to be 0.66 and 0.35 photons per shot in February and September of 2018, respectively.  The observed rates are therefore well below the theoretical expectations of roughly 12 and 5 photons per shot.  The ratios are similar, at 5.2\% and 6.7\% for the respective epochs.  It should be noted that the derating factors produce expectations roughly a factor-of-two different for the two epochs, and that the observed rates are likewise different by approximately the same factor---indicating consistent results.

We therefore conclude that observed return rates are at least an order-of-magnitude lower than expectations, and are unable to account for the difference by careful analysis and experimental characterization.  A proposed explanation that runs through this paper, and others before it \cite{Murphy:2010CCR,Murphy:2013Eclipse}, is that dust accumulation on the front surfaces of the CCRs simultaneously accounts for the link deficit, the additional full-moon deficit, and eclipse response.

Treating the dust as a set of geometrical obstructions, we recognize that the double-pass nature of the CCR makes each dust grain count \textit{twice}, blocking both entering and exiting light paths.  If the geometrical covering fraction is $f$, the probability of passage through the surface is $1-f$.  Independent probabilities for blockage in the two directions means that the probability of round-trip passage is $(1-f)^2$.  This amount of blockage will impact the central intensity of the far-field diffraction pattern (FFDP).  Normally, the unobstructed CCR, with front surface area, $A$, will have a FFDP central intensity proportional to $A^2$ (effectively, a coherent sum of the electric field over the open aperture, squared for intensity).  Since the obstructed CCR has an effective area of $(1-f)^2 A$, the FFDP central intensity will be reduced by $(1-f)^4$.  This strong dependence on $f$ means that, for instance, a 50\% covering fraction results in a sixteen-fold reduction in the FFDP central intensity---which is indeed approximately the factor of signal deficit we deduce.  Applied to the signal deficit ratios from the two epochs above, we compute dust fractions of 0.52 and 0.49 for February and September 2018, respectively.  The two epochs are therefore telling a very similar story in the context of dust coverage.

We note that infrared observations may be less impacted by dust, as the particles may be less ``geometrical" at longer wavelengths.  This may be related to the success of recent infrared LLR operations, where reflector performance appears to be better  \citep{Grasse2017}.  It is also the case the thermal lensing effect explored more below is less impactful at longer wavelengths.

%% file: eclipse_observations.tex
\begin{figure}[tb]
    \centering
\includegraphics{ 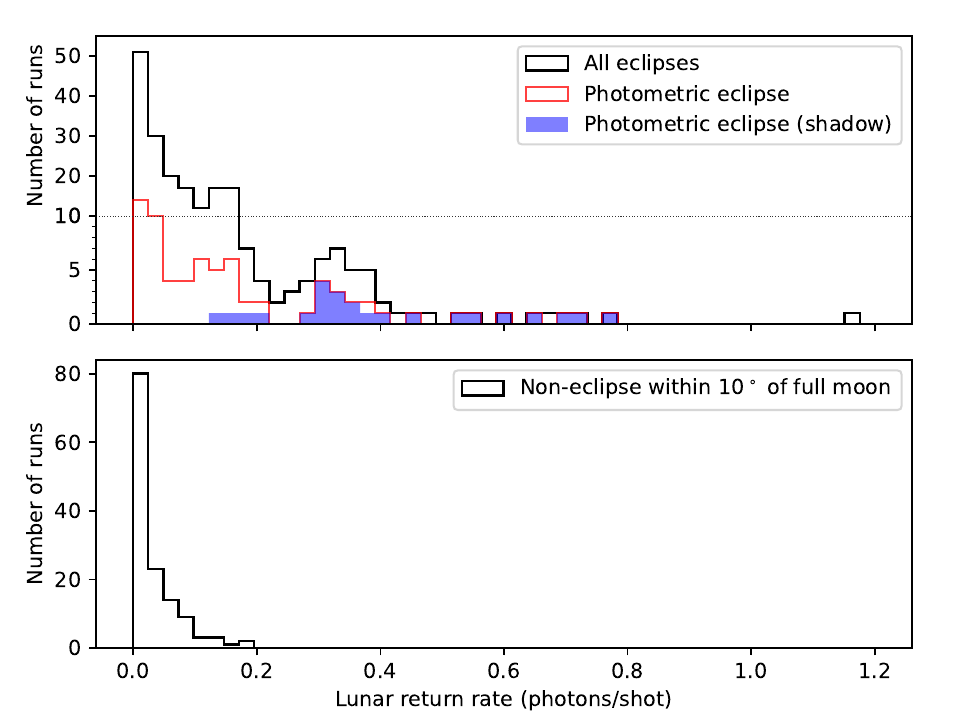}
    \caption{(Top) Histogram of the lunar return rate (photons per shot) during four eclipses (black line), as well as for the photometric eclipse on April 15, 2014 alone (red). Also shown (blue fill) is data from the photometric eclipse when the target reflectors are in full shadow. Note the change of scale on the ordinate indicated by the horizontal dotted line. (Bottom) Lunar return rate histogram when the moon is not in eclipse, but the lunar phase angle is close to full (within 10 degrees). All non-eclipse lunar return rates are less than 0.2 photons per shot.  Both histograms aggregate data from all three Apollo reflector arrays (Apollo 11, 14 and 15). The Apollo 11 and 14 rates have been scaled up by a factor of three because those arrays contain three times fewer corner cube reflectors than the Apollo 15 array.}
    \label{fig:rateHist}
\end{figure}

To study the effect of solar illumination on ranging throughput, we took advantage of the naturally occurring variation in the solar illumination incident on CCRs during lunar eclipses. During a lunar eclipse, the solar illumination changes from 100\% to 0\% (shadow) and back to 100\% over the course of about five hours as the CCRs enter and then exit the Earth's umbral shadow. A measurement of the evolution of the lunar return rate during the eclipse provides an independent constraint on the CCR dust coverage fraction, which can be compared to the fractions derived using link budget calculations (Section~\ref{sec:link-assessment}) and thermal eclipse simulations (later in Section~\ref{sec:simulation}).

We undertook four lunar eclipse ranging campaigns (December 21, 2010; April 15, 2014; September 28, 2015; January 21, 2019). Here, we focus on the 2014 eclipse (MJD 56762 from 5:52 to 10:52 UTC) because photometric sky conditions that night enabled the most reliable study of the throughput of the APOLLO system and lunar retroreflectors.
Range measurements were divided into short runs of $\sim$150 second duration (3000 laser shots) in order to monitor the evolving lunar return rate as a function of solar illumination of the lunar retroreflectors. 

We compare eclipse-night observations to lunar return rates measured during non-eclipse nights, but with the lunar phase within 10 degrees of full moon. In those cases, the median (mean) return rate is 0.02 (0.03) photons per shot, and the maximum is 0.2 photons per shot (see the bottom plot of Fig.~\ref{fig:rateHist}). 
During the 2014 eclipse, before the Apollo 11 and 14 reflectors entered the earth's umbral shadow the median (mean) return rate was also low, at 0.06 (0.08) photons per shot (red histogram in the upper plot of Fig.~\ref{fig:rateHist}).
After the reflectors entered the umbral shadow and began to cool (blue shaded histogram in the upper plot of Fig.~\ref{fig:rateHist}), the return rate rose as high as 0.77 photons per shot, with a median (mean) value of 0.35 (0.41) photons per shot, which is 17 times higher than the median rate during non-eclipse nights near full moon.
The lunar return rate histogram for data from all four eclipses is shown in black in the top plot of Fig.~\ref{fig:rateHist}, and it tells a consistent story: under overhead solar illumination, the optical throughput of the lunar reflectors degrades by at least an order of magnitude, but when the illumination is suppressed (via a lunar eclipse) and the reflectors cool, the rate increases. As described more fully in Section~\ref{sec:Eclipse}, we attribute the rate increase to the cooling of the reflectors when they are in shadow, which in turn reduces the thermal gradients in the CCR and improves the FFDP (and therefore the optical throughput) of the corner cubes.